# Nanoscale structural alterations in cancer cells to assess anti-cancerous drug effectiveness in cancer treatment using TEM imaging

Prakash Adhikari,[1] Mehedi Hasan,[1] Vijayalakshmi Sridhar,[2] Debarshi Roy,[3†] and Prabhakar Pradhan[1*]

[1]Department of Physics and Astronomy, Mississippi State University, Mississippi State, Mississippi 39762, USA
[2]Mayo Clinic College of Medicine, Department of Experimental Pathology, Minnesota, 39096 USA
[3]Department of Biology, Alcorn State University, Lorman, Mississippi 55905, USA

Emails: *PPradhan: pp838@msstate.edu ;
and  †DRoy: droy@alcorn.edu

**Abstract:** Understanding the nanoscale structural changes can provide the physical state of cells/tissues. It has been now shown that increases in nanoscale structural alterations are associated with the progress of carcinogenesis in most of the cancer cases, including early carcinogenesis. Anti-cancerous therapies are intended for the growth inhibition of cancer cells; however, it is challenging to detect the efficacy of such drugs in early stages of treatment. A unique method to assess the impact of anti-cancerous drugs on cancerous cells/tissues is to probe the nanoscale structural alterations. In this paper, we study the effect of different anti-cancerous drugs on ovarian tumorigenic cells, using their nanoscale structural alterations as a biomarker. Transmission electron microscopy (TEM) imaging on thin cell sections is performed to obtain their nanoscale structures. The degree of nanoscale structural alterations of tumorigenic cells and anti-cancerous drug treated tumorigenic cells are quantified by using the recently developed inverse participation ratio (IPR) technique. Results show an increase in the degree of nanoscale fluctuations in tumorigenic cells relative to non-tumorigenic cells; then a nearly reverse of the degree of fluctuation of tumorigenic cells to that of non-tumorigenic cells, after the anti-cancerous drugs treatment. These results support that the effect of anti-cancerous drugs in cancer treatment can be quantified by using the degree of nanoscale fluctuations of the cells via TEM imaging. Potential applications of the technique for cancer treatment are also discussed.

## I. Introduction:

***TEM imaging and probing nanoscale changes in cancer:*** TEM imaging is a method where we can probe ~1 nm resolution within the sample, and this has been used for imaging of cells at the nanoscale to see the inner structures of the cells, in general, qualitatively. It has been established that the cancer progression is associated with the nanoscale structural alteration in a cell due to the rearrangements of the building blocks of the cell such as DNA, RNA, and lipids. Therefore, TEM imaging is a good modality to probe cancerous changes in cells at the nanoscale level. Furthermore, the recently developed light wave localization technique, 'inverse participation ratio' (IPR), has shown success in quantifying the degree of structural changes in one parameter, known as the degree of structural disorder [1,2]. In particular, using the IPR method, a TEM image is used to construct a disordered 2D mass matrix, and from this we generate a 2D refractive index matrix. Optical waves are then solved for their eigenvalues and eigenfunctions using the refractive index matrix with closed boundary conditions. The light localization properties are measured by the average inverse participation ratio, *<IPR>*, and standard deviation of the IPR, *σ(IPR)*, of the eigenfunctions of the light waves in these samples. It is shown that the degree of structural disorder is proportional to the *<IPR>* or *σ(IPR)* [3,4]. Therefore, the *<IPR>* can be used as a measure of the degree of nanoscale structural disorder, and to monitor structural change in cells under diseases condition. The IPR method is a very versatile approach. The IPR method, using TEM imaging, has been recently generalized to study the structural changes in brain and colon cells in chronic alcoholism [5,6]. Furthermore, IPR method also extended to study the molecular specific (DNA, histone, etc.) structural changes in cells by using molecular specific fluorophores and confocal microscopy imaging [7,8].

*Ovarian cancer:*

Ovarian cancer (OC) ranks the 5$^{th}$ in cancer related deaths among women and accounts for more deaths than any other cancer of female reproductive system. The American Cancer Society (ACS), estimated new cases of OC in USA in 2019 would be 22,530, whereas estimated deaths would be 14,000. Most OC cases are diagnosed at a very late stage, of which 51% are diagnosed as stage III and 29% are diagnosed as stage IV [9]. The exact cause that triggers OC is not clearly understood but there are several risk factors such as fertility therapy, late pregnancy, family history, hormone therapy after menopause etc. are associated with the development of OC. Metabolic alterations, suppression of tumor suppressor genes, oncogenic activations are also considered as triggering factors for OC initiation and progression of disease [10,11]. Although initially sensitive to chemotherapy treatment, however majority of the OC patients develop chemo resistant. 10 years survival rate for most patients of all stages of OC is ~30%. Development of chemoresistance, widespread disease during the time of diagnosis and tumor recurrence are the major challenges in the therapeutics of ovarian cancer [12].

In this study, we are focusing on analyzing the impact of novel anti-cancerous drug treatment in the tumor forming OC cell line *in vitro*. HSulf-1 knockdown OV202 cells are selected for this study for their aggressive tumor forming ability and high proliferation rate [13,14]. This method of analysis is aimed to understand the effect of anti-cancerous drugs on cells in the early phases of treatment. Here we propose a novel approach to assess the impact of anti-cancerous drugs in cancerous cells by quantifying the degree of nanoscale structural disorder.

## II. Method:

*Analytical formulation for the inverse participation ratio (IPR) technique from TEM images:* TEM experiment has a resolution of ~1nm and can identify the nanoscale architectural alterations inside the cells which take place in normal cells when affected by cancer. These nano-alterations happen in the cells due to the rearrangement of the basic building blocks of the cells, such as DNA, RNA, lipids, macromolecules, etc. This results in mass density fluctuations in the cells. Using a thin slice of cell (~100nm), the mass density variations can be probed by TEM imaging. The IPR calculation is an efficient technique to measure and quantify the cancerous level of aggressiveness in a cell through its mass density fluctuations. A higher *<IPR> or σ(IPR)* value indicates an increasing amount of the nanoscale mass density fluctuations in cells. The IPR technique is described in detail in earlier publications [1,2,4,5]. However, for a self-sufficient and completeness of this paper, we will describe the IPR technique in brief.

The refractive index of a thin cell slice at a point *n(x,y)* with constant width *dz* has a voxel of volume *dV=dxdydz* which can be written as $n(x,y) = n_o + dn(x,y)$, where $n_o$ is the average refractive index and *dn(x,y)* is the fluctuation of refractive index at *(x,y)* indicated voxel.

TEM image intensity at any voxel point *(x,y)* for a thin cell sample is represented as $I_{TEM}(x,y)$ and can be expressed as $I_{TEM}(x,y) = I_{0TEM} + dI_{TEM}(x,y)$, where $I_{0TEM}$ is the average pixel intensity and $dI_{TEM}(x,y)$ is the fluctuation part of the pixel intensity. Here, the intensity fluctuation $I_{TEM}(x,y)$ is less than the average intensity $I_{0TEM}$ and similarly, the fluctuation of refractive index *dn(x,y)* is less than the average refractive index $n_0$.

Optical parameter refractive index *n(x,y)* of the scattering substances is linearly proportional to the mass density of a biological cell for the thin samples [1-2]. Therefore, the intensity of a TEM images is linearly proportional to the mass, M, and refractive index of the voxel:

$$I_{TEM}(x,y) \propto M(x,y) \propto n(x,y) \qquad (1a)$$
$$I_{0TEM} + dI_{TEM}(x,y) \propto M_0 + dM(x,y) \propto n_0 + dn(x,y) \qquad (1b)$$

From this, we can calculate optical potential of the voxel point as $\varepsilon_i(x,y)$ to generate an optical lattice:

$$\varepsilon_i = dn(x,y)/n_0 \propto dI_{TEM}/I_0 \quad (2)$$

Knowing the optical potential at every point, an Anderson disorder tight binding model TBM Hamiltonian [15-17] can be generated as follows:

$$H = \sum_i \varepsilon_i |i><j| + t\sum_{\langle ij \rangle}(|i><j| + |j><i|). \quad (3)$$

Where $|i>$ and $|j>$ are the eigenvectors of *i*-th and *j*-th lattice sites, $\varepsilon_i(x,y)$ or simply $\varepsilon_i$, is the *i*-th lattice site optical potential energy and *t* is the overlap integral between sites *i* and *j*. The eigenfunctions ($E_i$'s) can now be generated from the above Hamiltonian by its diagonalization. Finally, the average IPR value of the whole samples we can define as [1-4, 18,19]:

$$<IPR>_N = \frac{1}{N}\sum_{i=1}^{N} \int_0^L \int_0^L E_i^4(x,y)dxdy, \quad (4)$$

where $E_i$ denotes the *i*-th eigenfunction of the Hamiltonian, *N* is the total number of potential points on the refractive index matrix (i.e., $N=(L/dx)^2$). It has been shown that the average IPR value, *<IPR>*, is proportional to the degree of structural disorder $L_d=dn \times l_c$, where *dn* is the *std* of the all *n(x,y)* point and $l_c$ is the spatial correlation decay length of the *n(x,y)* over the sample [1,2]. Then,

$$\langle IPR \rangle \equiv <IPR>_{ensemble} \sim L_d = dn \times l_c, \quad (5a)$$

$$\sigma(IPR) \equiv \sigma(IPR)_{ensemble} \sim L_d = dn \times l_c. \quad (5b)$$

Our statistical analysis involves calculating the average and standard deviation of the disorder strength of IPR values, i.e. $L_d$ values over the samples. Using this structural disorder strength *<IPR> or σ(IPR)* or $L_d$ as a biomarker, we study the structural properties of the ovarian cancer cells with anti-cancerous drug treatments. In particular, we expect an increase in the structural disorder with the growth of cancer, and the reversibility of the structural disorder when the cancerous cells are treated with anti-cancerous drug, if the nanoscale structural disorder is a good cancer stage biomarker.

### III. Sample Preparation and TEM imaging

*Ovarian normal and cancer cell lines:* OV202 cell line is a low-passage primary ovarian cancer cell line established at the Mayo Clinic [13]. OV202 NTC (expressing HSulf-1) and Sh1 cells (HSulf-1 deficient) are developed by Dr. Shridhar's group at Mayo Clinic and is described earlier elsewhere [14]. Subcutaneous injection of OV202 Sh1 cells resulted in tumor formation in nude mice, whereas HSulf-1 expressing OV202 NTC cells did not form tumor [13]. Both cells were grown in minimum essential medium alpha 1X (Cellgro) supplemented with 20% fetal bovine serum (Biowest) and 1% penicillin-streptomycin (Cellgro). All cells were grown in the presence of 1 μg/ml puromycin as a selection marker for the HSulf-1 shRNA cells were treated with 10 µl of AACOCF3 or MAFP (cPLA2 inhibitors; Cayman chemicals) for 24 hours. Following this treatment, cells were washed twice with PBS and then fixed in Trump's fixative containing 4% formaldehyde and 1% glutaraldehyde in a phosphate buffer pH ~7.3, post-fixed in 1.0% OsO4, dehydrated with ethanol gradation, and transitioned into propylene oxide for infiltration and embedding into super epoxy resin.

***TEM imaging:*** Cell samples were fixed in Trump's fixative (pH 7.2) at 4°C overnight, spun down and the supernatant removed. They were re-suspended in agarose which was cooled and solidified. The cells in agarose were then post-fixed in 1% OsO4, dehydrated through a graded series of ethanol and embedded in Spur resin. 100nm (or 0.1 μm) ultra-thin sections were mounted on 200-mesh copper grids, post-stained with lead citrate, and observed under a JEOL JEM-1400 transmission electron microscope at 80kV.

**IV. Results and discussions:**

The TEM images of the ovarian cancer cells are obtained as described in the above section. IPR analyses were performed for the samples on *165nm×165nm* TEM images. IPR averaging over the single cells, as well as over the different cells, were performed for ensemble averaging. As discussed above, the average <*IPR*> and σ(IPR) value for each TEM image was calculated and provides the degree of the structural disorder strength at a defined length scale.

Figure 1(a)-(d) are the representative grayscale TEM images of a thin section (~100nm) of cell from the following ovarian cell lines: (i) non-tumorous NTC, (ii) tumorous Sh1, (iii) Sh1 treated with drug AACOCF3 (Sh1-AACOCF3), and (iv) Sh1 treated with drug MAPF (Sh1-MAFP). For each case study, ~ 8-10 different cells TEM images were taken from the cell line for averaging. Figure 1(a')-(d') are the corresponding σ*(IPR)* images of Figure 1(a)-(d), at a length scale of 165nm and sample size of *165×165nm$^2$*. As can be seen from Figure 1, σ(*IPR)* images represent different intensities of disorder pattern in the cell line distinctly than conventional grayscale TEM images. In the IPR images, intensities

patterns of higher fluctuations in the cells are represented by the red spots and lower intensities with blue. In the figure, it can be seen that the increasing fluctuation or $\sigma(IPR)$ value increases from the less proliferating NTC cells to highly proliferating Sh1cells, and decreasing of the fluctuations or $\sigma(IPR)$ values decreases with the treatment of two different anti-cancerous drugs, AACOCF3 and MAFP. The drug effect can be distinctly visualized in the IPR images.

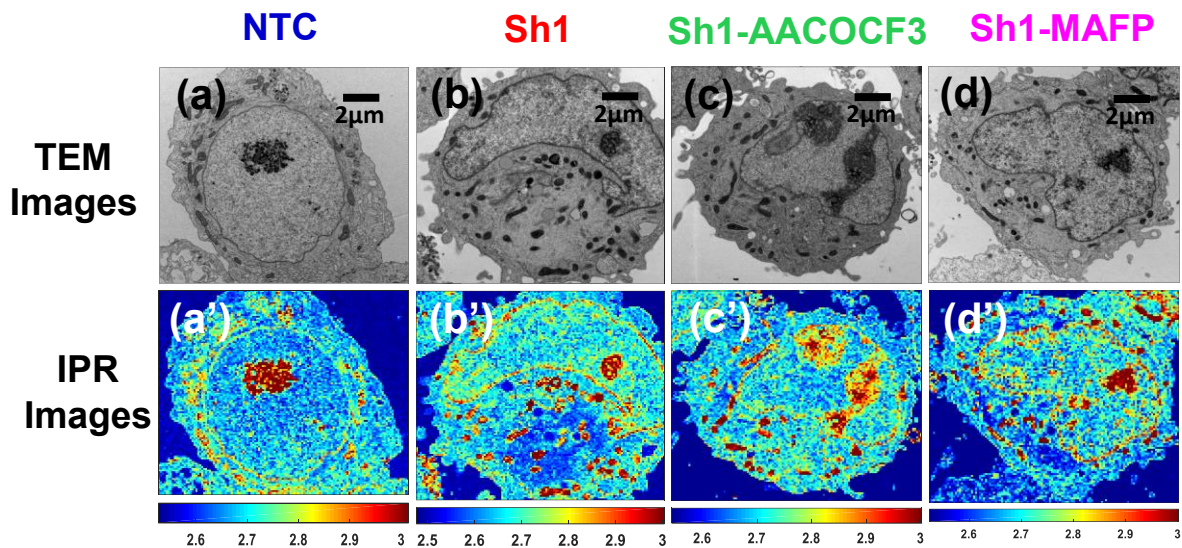

**Fig 1: (a)-(d)** are the representative TEM images and **(a')-(d')** are their respective IPR images from ovarian cells of the following: non-tumorous (OV202 NTC); tumorous (OV202 Sh1); AACOCF3 treated tumorous Sh1, Sh-AACOCF3; and MAFP treated tumorous Sh1 Sh1-MAFP. IPR images are distinct from the TEM images.

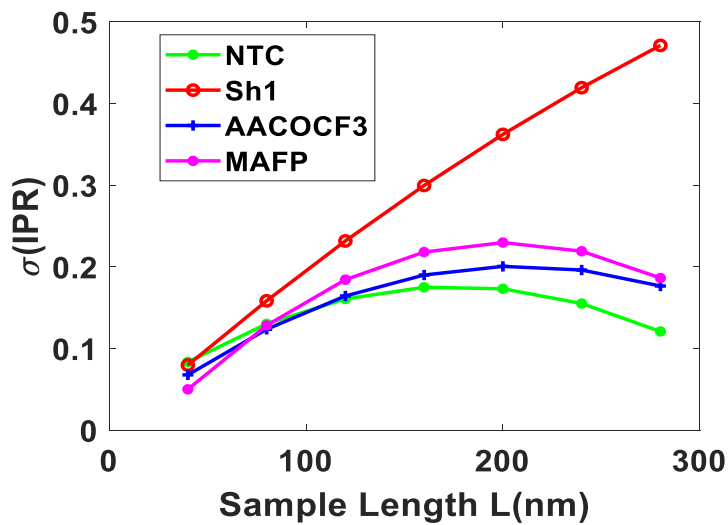

**Fig. 2:** Variations of the standard deviation σ(IPR(L)) with the increase of length scale L, for cell lines non-tumorigenic (OV202 NTC); tumorigenic (OV202 Sh1); AACOCF3 treated tumorigenic Sh1 (Sh1-AACOCF3); and MAFP treated tumorous Sh1 (Sh1-MAFP). The ensemble averagins were performed over TEM images of ~8-10 different cells. It can be seen that the deviation between NTC and Sh1 started becoming prominent around the length scale ~100 nm. Interestingly, the drug treated Sh1 tumorigenic cells fluctuation degrees reverse to the non-tumorigenic cells.

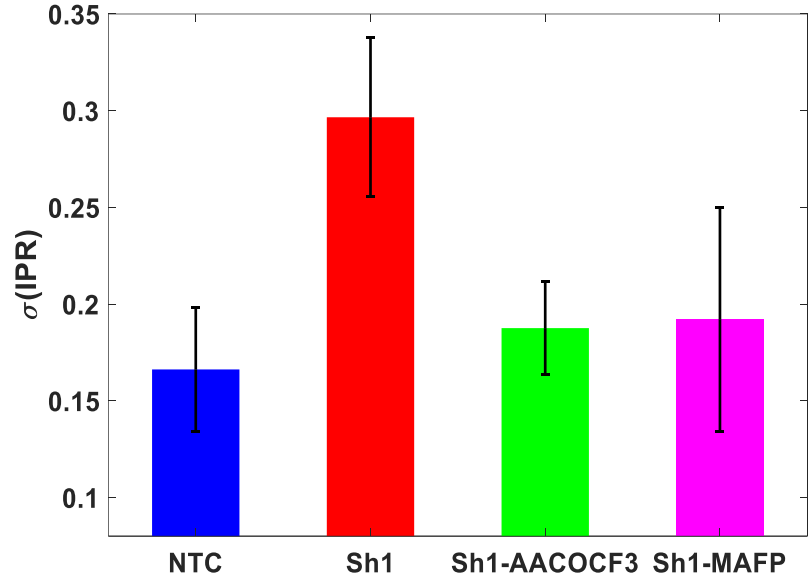

**Fig. 3:** Bar graph representation of the standard deviation of the degree of structural disorder strength $L_d \sim \sigma(IPR)$ calculated from the TEM images for: non-tumorous (NTC), tumorous (Sh1), AACOCF3 treated tumorous (Sh1) cells Sh1-AACOCF3, and MAFP treated tumorous cells (Sh1) Sh1-MAFP. The normal cells,-IPR analysis was performed at the sample size $165 \times 165\ nm^2$. The result shows $L_d$ value increases from non-tumorous to tumorous cells, then it decreases when these tumorous cells are further treated with anti-cancerous drugs AACOCF3 and MAFP, interestingly the $L_d$ value returns almost back to the same value of the non-tumorous cells (p-value < .05). This may imply that anti-cancerous drugs are working well in ovarian cancer treatment.

Figure 2 shows length (*L*) dependent fluctuations with the sample size (L×L). We plotted, variations of the standard deviation $\sigma(L)$ with the increase of sample lengths: $L$ = 41, 82, 123, 165, 206, 247, 288 nm. These lengths are for the cells from the following cell lines: non-tumorigenic (OV202 NTC); tumorigenic (OV202 Sh1); AACOCF3 treated tumorigenic Sh1 (Sh1-AACOCF3); and MAFP treated tumorous Sh1 (Sh1-MAFP). As can be seen from the figure that the deviation in the degree of nano-fluctuations between non-tumorigenic cells NCT and tumorigenic cells Sh1 started becoming prominent around the length scale ~100 nm. Interestingly, the degree of nano-fluctuations of anti-cancerous drugs treated Sh1 tumorigenic cell reverse to that of the non-tumorigenic NTC cells. This confirms the efficacy of these two anti-cancerous drugs.

Figure 3 presents the bar graphs of the standard deviation of calculated $\sigma(IPR)$ or $(L_d)$ value, of the ovarian cells at the fixed length scale 165nm. The variations are similar at lower sample length scales >165nm, however we have chosen 165nm to show a prominent difference. Statistically, the standard deviation is the more reliable marker than the average, as it only depends on the width of the distribution, irrespective to the mean position. The result shows the standard deviation of the degree of structural disorder $\sigma(IPR)$ value increased by 70% from NTC to Sh1 cells. Furthermore, when Sh1 cells were treated with 2 different anti-cancerous drugs, AACOCF3 and MAFP, the $\sigma(IPR)$ values decreased by around 60% for AACOCF3 and 50% for MAFP, relative to the $\sigma(IPR)$ value of the Sh1 tumorous cells. In particular, with the treatment of the anti-cancerous drug, the structural biomarker parameter $\sigma(IPR)$ or $L_d$ value decreased nearly back to the normal value. The normalcy detection of these anti-cancerous drug treated cancerous cells may require further investigations using different modalities. It has been earlier shown that AACOCF3 is a better anti-cancerous agent producing more anti-cancerous effects in OC cells compared to MAFP [20]. It can be seen in Fig. 3 that similar trend of bar graphs which show a reduction in the degree of structural disorder $\sigma(IPR)$ value for AACOCF3 (60%) > MAFP(50%), consistent with the known qualities of the drugs, in this length scale. Hence, the quantitative analysis technique, called IPR, quantifies the nanoscale structural disorder $\sigma(IPR)$ or $L_d$, as an important biomarker to study the structural alterations at nanoscale level and has potential to detect the effect of anti-cancerous drugs in carcinogenesis ovarian cancer.

## IV. Conclusions

The nanoscale mass-density fluctuations are quantified with the progression of carcinogenesis, as wells as the effects of two anti-cancerous drugs on non-tumor forming OV202NTC and tumor forming OV202Sh1 cells are studied using TEM imaging and IPR technique. The nanoscale fluctuations are quantified by the *std* value of the *IPR*, $\sigma(IPR)$, performed over an ensemble of samples. Results show an increase in the nanoscale fluctuations or σ*(IPR)* value from non-tumorous NTC to tumorous Sh1 cells. The $\sigma(IPR)$ values for two different drugs treated tumorous cells, Sh1-AACOCF3 and Sh1-MAFP, have reduced value of the $\sigma(IPR)$ from tumorous cells Sh1 and the reduced values are nearly same to the NTC non-tumorous cells. Earlier IPR analysis of a different cell line has verified the increase of nanoscale structural disorder with the progression of cancer [1,2]. Based on the results presented, we investigate the potential applications of the IPR technique in measuring and quantifying the effectiveness of different anti-cancerous drugs on ovarian cancer treatment. This quantification of effectiveness of anti-cancerous drugs in ovarian cancer treatment could enhance better drugs treatment modalities at its earliest and helps to control the deadly ovarian cancer. Although this study is based on ovarian cancer, however, the technique can be applied to the varieties of cancers to assess the effectiveness of different anti-cancerous drugs in treatment.

**Acknowledgements**

This work was partially supported by: NIH R01EB016983 and Mississippi State University (PP); Mississippi INBRE, an Institutional Development Award (IDeA), NIH P20GM103476 (DR); NIH CA106954 and the Department of Experimental Pathology and Laboratory Medicine and the Mayo Clinic (VS). Special thanks to Scott Gamb for helping in TEM imaging in Microscopy and Cell Analysis core at Mayo clinic.